# Applied Symbolic Vector Dynamics of Coupled Map Lattice

Kai Wang, and Wenjiang Pei

Department of Radio Engineering, Southeast University, Nanjing, China

Symbolic dynamics, which partitions an infinite number of finite-length trajectories into a finite number of trajectory sets, describes the dynamics of a system in a simplified and "coarse-grained" way with a limited number of symbols. The study of symbolic dynamics in 1D chaotic map has been further developed and is named as the applied symbolic dynamics. In this paper, we will study the applied symbolic vector dynamics of CML. Based on the original contribution proposed in Refs.[6], we will study the ergodic property of CML. We will analyze the relation between admissibility condition and control parameters, and then give a coupling coefficient estimation method based on the ergodic property. Both theoretical and experimental results show that we provide a natural analytical technique for understanding turbulences in CML. Many of our findings could be expanded to a wider range of application.

Symbolic dynamics, which partitions an infinite number of finite-length trajectories into a finite number of trajectory sets, describes the dynamics of a system in a simplified and "coarse-grained" way with a limited number of symbols[1]. Today, the study of symbolic dynamics in 1D chaotic map has been further developed and is named as the applied symbolic dynamics. In Ref.[1], the applied symbolic dynamics mainly focuses on the following four aspects: 1) the relation between symbols and monotonic inverse chaotic functions; 2) the relation between symbolic sequences and chaotic sequences; 3) the relation between symbolic sequences and chaotic control parameters; 4) symbolic description of the ergodic property. The effective symbolic description connects the dynamical system theory with the study of communication, signal estimation, cryptology, etc.[2]. Coupled map lattice (CML), which utilizes a coarse space-time discretization and continuous state variable, can reproduce the essential features of spatiotemporal phenomena and has been investigated as theoretical models in a variety of researches, covering physics, neuroscience, and computer engineering [3]. Researches in symbolic dynamics have been expanded to spatio-temporalized and networklized areas, and thus the study of CML has become a rapidly growing part of symbolic dynamics[4-7]. There is one-to-one correspondence between the set of global orbits and set of admissible codes. As a result, CML with $N$ sites can be fully described by $2^N$ symbolic vectors.

In this paper, we will study the applied symbolic vector dynamics of CML. Based on the original contribution proposed in Refs.[6], we will study the ergodic property of CML. We will analyze the relation between admissibility condition and control parameters, and then give a coupling coefficient estimation method based on the ergodic property. Both theoretical and experimental results show that we provide a natural analytical technique for understanding turbulences in CML. Many of our findings could be expanded to a wider range of application.

Let's consider the typical CML with $N$ sites labeled. Each site is described by a state $x_n^i$ in the interval $I = [a,b]$ and has local dynamics $f_i : I \to I$. $e$ is the coupling coefficient, $n = 1, 2, \cdots$ is the time index, $i = 1, 2, \cdots N$ is the lattice site index and $f_i$ is unimodal map such as Logistic map $f_i(x) = 1 - 2x^2$ [5,6].

$$x_{n+1}^i = (1-e)f_i(x_n^i) + e/2[f_{i-1}(x_n^{i-1}) + f_{i+1}(x_n^{i+1})] \tag{1}$$

Denote $F$ as the product function of $f_i$ onto each site, i.e., $F(\mathbf{x}_n) = \mathbf{fx}_{n+1}$, where $\mathbf{x}_n = [x_n^1, \cdots, x_n^N]^T$, $\mathbf{fx}_{n+1} = [f_1(x_n^1), \cdots, f_N(x_n^N)]^T$. Then Eq. (1) can be generalized as the composition $\mathbf{x}_{n+1} = A \times \mathbf{fx}_{n+1} = A \circ F(\mathbf{x}_n)$, where $A$ is the coupling matrix of Eq. (1). Let $H = A \circ F$, $H : I^N \to I^N$, and then Eq. (1) can be generalized as $\mathbf{x}_{n+1} = H(\mathbf{x}_n)$ [5].

Symbolic vector dynamics gives a strict correspondence between the set of global orbits and the set of admissible codes. Let's label location of the maxima as the critical point $x_c^i$ and divide the interval $I = [a, b]$ into two sets $D^i = \{d_{-1}^i, d_1^i\}$, where $d_{-1}^i = [a, x_c^i]$ and $d_1^i = (x_c^i, b]$. CML can generate symbolic vector sequence $\mathbf{S} = \{\mathbf{s}_0, \mathbf{s}_1, \cdots, \mathbf{s}_n, \cdots\}$, where $\mathbf{s}_n = [s_n^1, \cdots, s_n^N]^T$ at the $n$th iteration and

$$s_n^i = \begin{cases} -1 & \text{if } x_n^i \in d_{-1}^i \\ 1 & \text{if } x_n^i \in d_1^i \end{cases}$$

Let $\mathbf{fx}_{n+1} = A^{-1} \times \mathbf{x}_{n+1}$ and $fx_{n+1}^i = f_i(x_n^i)$, where $\mathbf{fx}_{n+1} = [f_1(x_n^1), \cdots, f_N(x_n^N)]^T$. When symbol $s_n^i$ is known, the function $f_{i\ s_n^i}^{-1}(fx_{n+1}^i) = x_n^i$ is one-to-one correspondent, because $f_i$ is a monotonic function in every interval $d_m^i$. Denote $F_{\mathbf{s}_n}^{-1}$ as the product function of $f_{i\ s^i}^{-1}$ at the $i$th site and the $n$th iteration when $s_n^i$ is known, and then $F_{\mathbf{s}_n}^{-1}(\mathbf{fx}_{n+1}) = \mathbf{x}_n$, where $\mathbf{x}_n = [f_{1\ s_n^1}^{-1}(fx_{n+1}^1), \cdots, f_{N\ s_n^N}^{-1}(fx_{n+1}^N)]^T$. That is, when $\mathbf{s}_n$ is known, the inverse function $F_{\mathbf{s}_n}^{-1} \circ A^{-1}(\mathbf{x}_{n+1}) = \mathbf{x}_n$ can be confirmed unambiguously. Define $H_{\mathbf{s}_n}^{-1} = F_{\mathbf{s}_n}^{-1} \circ A^{-1}$, $H_{\mathbf{s}_n}^{-1} : I^N \to I^N$, and then the inverse function of CML can be generalized as $\mathbf{x}_n = H_{\mathbf{s}_n}^{-1}(\mathbf{x}_{n+1})$ unambiguously[6].

Although the inverse function $f_s^{-1}$ is ergodic in $I$, the function $H$ of CML is not ergodic in $I^N$. Let $B = \{b_{ij}\}_{i,j=1,2\cdots N}$ be the inverse matrix of the coupling matrix $A$. For any vector $\mathbf{x}' = [x'_1, x'_2, \cdots, x'_N]^T \in I^N$, in order to calculate $H_{\mathbf{s}_n}^{-1}(\mathbf{x}')$, we should calculate $B \times \mathbf{x}' = \mathbf{y}' = [y'_1, y'_2, \cdots, y'_N]^T$ first, where $y'_j = \sum_{j=1}^{N} b_{ij} x'_i$, $j = 1, 2, \cdots, N$. Consider the inverse functions of Logistic map $f_{i\ s^i}^{-1}(u) = s^i \sqrt{c_i - u}$. When $u > 1$, the value of $f_s^{-1}(u)$ is a complex number; when $u < -1$, $f_s^{-1}(u) \notin I = [-1,1]$. As a result, for any vector $\mathbf{x}' = [x'_1, x'_2, \cdots, x'_N]^T \in I^N$, if its forward value in one step $H_s^{-1}(\mathbf{x}') \notin I^N$ with any symbolic vector $\mathbf{s}$, this vector $\mathbf{x}'$ is a forbidden word and can not be generated by iteration of CML.

Select an initial vector value randomly, and then iterate CML ($N$=2) 1000 steps. As shown in Fig.1(a), the red points describe the dynamical trajectory. Those points are distributed in a diamond-like region surrounded by four black beelines. Any $\mathbf{x}_i$ out of this diamond-like region is a forbidden word, because its forward value $H_{\mathbf{s}_i}^{-1}(\mathbf{x}_i) \notin I^N$ and can not be generated by iteration. However, as shown in Fig.1(a), we can not describe the forbidden word region exactly by using the four black beelines. In other words, any vector $\mathbf{x}'$ which satisfies $H_s^{-1}(\mathbf{x}') \notin I^N$ must be a forbidden word, but not every vector $\mathbf{x}'$ which satisfies $H_s^{-1}(\mathbf{x}') \in I^N$ can be generated by iteration. Furthermore, consider the forward value in two steps of $\mathbf{x}'$: $H_s^{-1}H_s^{-1}(\mathbf{x}')$. Any vector $\mathbf{x}'$, which satisfies $H_s^{-1}H_s^{-1}(\mathbf{x}') \notin I^N$ with any symbolic vector sequence $\{\mathbf{s},\mathbf{s}\}$, is still a forbidden word. In Fig.1(b), the blue part is the forbidden word region $K^N$. Any vector $\mathbf{x}' \in K^N$ satisfies $H_s^{-1}H_s^{-1}(\mathbf{x}') \notin I^N$ with any symbolic vector sequence $\{\mathbf{s},\mathbf{s}\}$. Obviously, this region can describe the forbidden word region better.

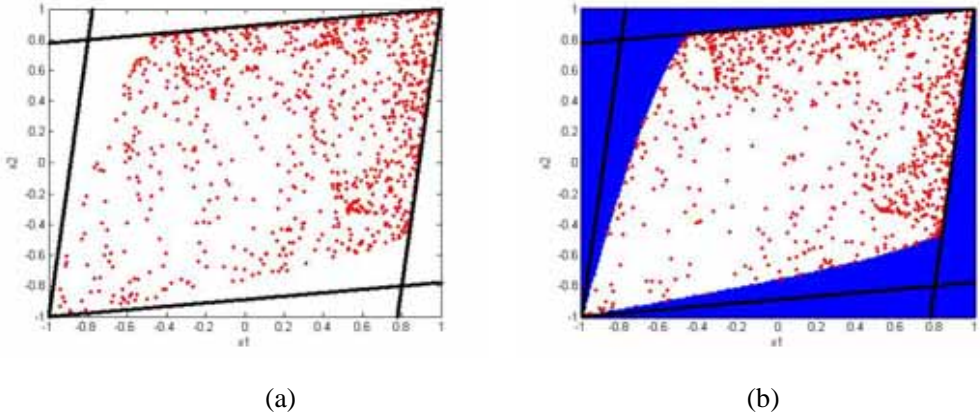

(a)          (b)

Fig.1 The ergodic property of CML ($N$=2). The functions of the four black beelines are $b_{11}x_1 + b_{12}x_2 = \pm 1$ and $b_{12}x_1 + b_{11}x_2 = \pm 1$ respectively, where $b_{ij}$ is the element of a matrix $B$ in the $i$th row and the $j$th column. The red points describe the dynamical trajectory and the blue part is the forbidden word region $K^N$.

The symbolic vector sequence $\{\mathbf{s}_n, \mathbf{s}_{n+1}, \cdots, \mathbf{s}_{n+L}\}$ is equivalent to the sub-region $J^N$ of $I^N$. When the length of symbolic vector sequence $L \to \infty$, the sub-region $J^N$ will converge to an vector with real values. According to the original contribution proposed in Refs.[6-7], Eq.2 is held.

$$\lim_{L \to \infty} \mathbf{x}(n \mid L) = \mathbf{x}(n) \quad (2)$$

where $\mathbf{x}(n \mid L) = H^{-L}_{\mathbf{s}_n, \cdots, \mathbf{s}_{n+L-1}}(\boldsymbol{\eta})$, $\boldsymbol{\eta}$ is any vector in $I^N$ and $\mathbf{x}(n)$ is the initial vector of the symbolic vector sequence $\{\mathbf{s}_n, \mathbf{s}_{n+1}, \cdots, \mathbf{s}_{n+L}\}$. If the corresponding sub-region $J^N$ belongs to the forbidden word region $K^N$, then the symbolic vector sequence $\{\mathbf{s}_n, \mathbf{s}_{n+1}, \cdots, \mathbf{s}_{n+L}\}$ is a forbidden word. For any symbolic vector sequence $\{\mathbf{s}_i\}_{i=1}^M$, let region $J^N$ be the mapping of the region $I^N$ according to the rule $H^{-M}_{\{\mathbf{s}_i\}_{i=1}^M}$. We can judge whether the symbolic vector sequence $\{\mathbf{s}_i\}_{i=1}^M$ is a forbidden word through the location of $J^N$.

Let's take the Logistic map based CML with 2 sites as an example. Encode the symbolic vector $(s_n^1, s_n^2)^T$ as the symbol $S_n \in \{0, 1, 2, 3\}$:

$$\begin{cases} S_n = 0 & \text{when} \quad (s_n^1, s_n^2)^T = (-1, -1)^T \\ S_n = 1 & \text{when} \quad (s_n^1, s_n^2)^T = (1, -1)^T \\ S_n = 2 & \text{when} \quad (s_n^1, s_n^2)^T = (-1, 1)^T \\ S_n = 3 & \text{when} \quad (s_n^1, s_n^2)^T = (1, 1)^T \end{cases}$$

The location of $J^2$ according to the rule $H^{-3}_{\{S_i\}_{i=1}^3}$ is shown in Fig.2. When $J^2 \subseteq K^2$, the corresponding symbolic sequence $\{S_i\}_{i=1}^3$ can be judged as a forbidden word. The forbidden words $\{S_i\}_{i=1}^3$ in different coupling coefficients are shown in Tab.1. In Ref.[5], a different conclusion of the forbidden words $\{S_i\}_{i=1}^3$ is given. When $\varepsilon = 0.1$, the forbidden words $\{S_1, S_2, S_3\}$ are 010, 020, 100, 101, 102, 110, 111, 120, 122, 200, 201, 202, 210, 211, 220, 222, 310, 320[]. However, as shown in Fig.2, the conclusion is not accurate and needs further improvement.

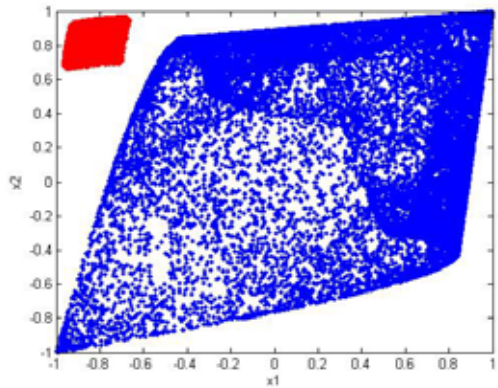

Fig.2 The red part is the corresponding region $J^2$ of the symbolic sequence $\{S_i\}_{i=1}^3$. The symbolic sequence {2 0 3} is not judged as a forbidden word in Ref.[5]. However, as shown in Fig.2(b), $J^2 \subseteq K^2$. As a result, {2 0 3} must be a forbidden word.

Tab1: Forbidden 3-block words with different coupling coefficients

| | $\{S_1, S_2, S_3\}$ | | | | | | | | | | | | |
|---|---|---|---|---|---|---|---|---|---|---|---|---|---|
| $\varepsilon = 0.1$ | 010 | 011 | 020 | 022 | 100 | 101 | 102 | 103 | 110 | 111 | 113 | 120 | 122 | 200 |
| | 201 | 202 | 203 | 210 | 211 | 220 | 222 | 223 | 310 | 320 | | | | |
| $\varepsilon = 0.08$ | 010 | 011 | 020 | 022 | 100 | 101 | 102 | 103 | 110 | 111 | | 120 | 122 | 200 |
| | 201 | 202 | 203 | 210 | 211 | 220 | 222 | | 310 | 320 | | | | |
| $\varepsilon = 0.06$ | 010 | | 020 | | 100 | 101 | 102 | | 110 | 111 | | 120 | 122 | 200 |
| | 201 | 202 | | 210 | 211 | 220 | 222 | | | | | | | |
| $\varepsilon = 0.04$ | | | | | 100 | 101 | 102 | | 110 | 111 | | | | 200 |
| | | 202 | | | | | 222 | | | | | | | |

Different dynamical systems have different dynamics. Dynamical characteristics of CML are determined by the coupling coefficient, when local dynamics $f_i : I \to I$ is determined. There is one-to-one correspondence between the coupling coefficient and the symbolic vector sequence. For the Skewed map based CML, we can estimate the coupling coefficient from the symbolic vector sequence by using the *word lifting* method when the initial value is known[6]. As shown in Tab.1, there is correspondence between the forbidden word and the coupling coefficient. The total number of forbidden words is increasing when the coupling coefficient is increasing. Different coupling coefficients have different sets of forbidden words. Let's also take the Logistic map based CML with 2 sites as an example. Suppose the coupling coefficient is $\varepsilon_1$. The corresponding forbidden word interval is $K^2(\varepsilon_1)$. CML generates the dynamical sequence $\mathbf{X}(\varepsilon_1)$ and symbolic vector sequence $\mathbf{S}(\varepsilon_1)$. According to Eq.(2), let's recover the dynamical sequence with the estimation value $\varepsilon_2$ by using $\mathbf{S}(\varepsilon_1)$, and let the estimation sequence be $\mathbf{X}(\varepsilon_2)$. Obviously, $\{\mathbf{x} | \mathbf{x} \in \mathbf{X}(\varepsilon_2) \bigcap \mathbf{x} \in K^2(\varepsilon_1)\} = \emptyset$ because $\mathbf{X}(\varepsilon_1) = \mathbf{X}(\varepsilon_2)$ when $\varepsilon_1 = \varepsilon_2$; when $\varepsilon_1 \neq \varepsilon_2$, $\{\mathbf{x} | \mathbf{x} \in \mathbf{X}(\varepsilon_2) \bigcap \mathbf{x} \in K^2(\varepsilon_1)\} \neq \emptyset$. Let the value which belongs to $\mathbf{X}(\varepsilon_2)$ and $K^2(\varepsilon_1)$ be a forbidden word. Select the initial value $\mathbf{x}_0 = [0.18, 0.04]^T$ randomly, and then generate $\mathbf{S}(\varepsilon_1)$ with the coupling coefficient $\varepsilon_1 = 0.51$. The relation between $\varepsilon_2$ and the total number of forbidden words is given in Fig.3. When $\varepsilon_1 = \varepsilon_2$, the total number of the forbidden word is 0. When $\varepsilon_2 < \varepsilon_1$, the total number is decreasing with the increase of $\varepsilon_2$. When $\varepsilon_2 > \varepsilon_1$, the total number is increasing with the increase of $\varepsilon_2$.

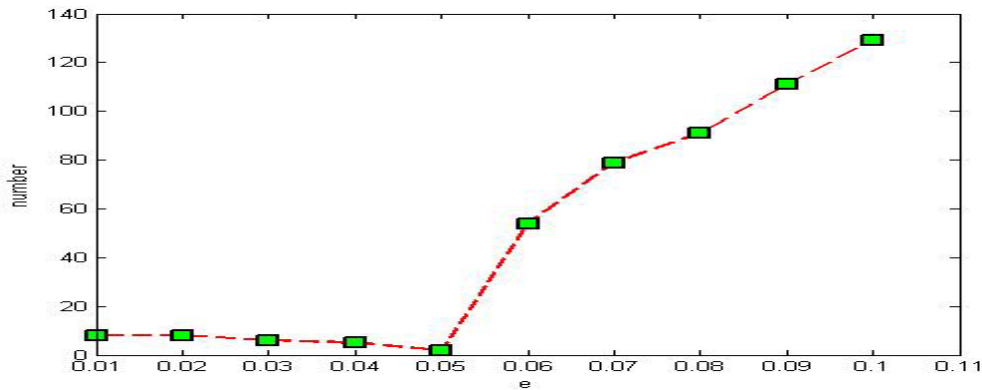

Fig.3 The relation between $\varepsilon_2$ and the total number of forbidden words $L=30$.

In this paper, we study the applied symbolic vector dynamics of CML. Based on the original contribution proposed in Ref.[6], we study the ergodic property of CML. We analyze the relation between admissibility condition and control parameters, and then give a coupling coefficient estimation method based on the ergodic property. Both theoretical and experimental results show that we provide a natural analytical technique for understanding turbulences in CML. Many of our findings could be expanded to a wider range of application.

This work was supported by the Natural Science Foundation of China under Grant No 60672095, the National High-Technology Project of China under Grant 2007AA11Z210, the Doctoral Fund of Ministry of Education of China under Grant 20070286004, and the Special Scientific Foundation for the "Eleventh-Five-Year" Plan of China.

References
[1] B.L.HaoL *Starting With Parabolas-An Introduction to Chaotic Dynamics* (Shanghai scientific and Technological Education Publishing House, Shanghai, 1994); W.M.Zheng and B.L.Hao *Applied Symbolic Dynamics* (Shanghai scientific and Technological Education Publishing House, Shanghai, 1994).
[2] S. Jorg and S. Thomas, *IEEE Trans. Circuits Syst. I* 48, 1269 (2001); S. Jorg and S. Thomas, *IEEE Trans. Circuits Syst. I*, 48, 1283 (2001); C. Ling, X.F. Wu and S.G. Sun, *IEEE. Trans. Signal Process*. 47, 1424 (1999); F. J. Escribano, L. Lopez andA. F. Sanjuan, *Electron. Lett.*, 42, 984 (2006); E.R.Jr., S. Hayes, C. Grebogi, *Phys. Rev. Lett.* 96, 1247 (1997).
[3] W.M.Yang 1994 *Spatiotemporal Chaos and Coupled Map Lattice* (Shanghai scientific and Technological Education Publishing House, Shanghai, 1994).
[4] W. Just, J. Stat. Phys. 90, 727 (1998); 105, 133 (2001); R. Coutinho and B.Femandez, Physica D 108, 60 (1997); Y.C.Zeng and Q.Y.Tong, *Acta Phys. Sin.* 52, 285 (2003); Y.Liu, M.F.Shen and F.H.Y. *Chan, Acta Phys. Sin.* .55, 564 (2006); S. D. Pethel and E. Bollt, *Phys. Rev. Lett.* 99, 214101 (2007).
[5] S. D. Pethel, N. J. Corron and E. Bollt, *Phys. Rev. Lett.* 96, 034105 (2006);
[6] K.Wang, W.J.Pei, S.P.Wang and Y.M. Cheung, *IEEE Trans. Circuits Syst. I* 55, 1116 (2008);
[7] K.Wang, W.J.Pei, H.S.Xia and Z.Y. He, *Acta Phys. Sin.*.56, 130 (2007); K.Wang, W.J.Pei, Z.Y.He and Y.M. Cheung , Phys. Lett. A 367, 316 (2007)